\documentclass{PoS}
\usepackage{axodraw}
\usepackage{epsfig}

\newcommand {\sla} {\slash \hspace{-0.22cm}}

\title{Wick Rotation in the Light-Front} %% Field Theory}
\ShortTitle{Wick Rotation in the Light-Front}

\author{\speaker{J.~P.~B.~C.~de~Melo$^a$}\thanks{lftcc.unicsul.com.br},
~E.~Gambim$^b$ and T.~Frederico$^b$\\
        $^a$Laborat\'orio de F\'\i sica Te\'orica e Computa\c{c}\~ao
Cient\'\i fica,
Universidade Cruzeiro do Sul, 08060-700  and
Instituto de F\'isica Te\'orica, 01405-900, S\~ao Paulo, Brazil \\
$^b$Dep. de F\'isica, Instituto
Tecnol\'ogico de Aeron\'autica, 12.228-900, S\~ao Jos\'e dos Campos, Brazil
        E-mail: \email{pacheco@ift.unesp.br}}
\abstract{We study the electroweak properties of pseudo-scalar
mesons in the light and heavy-light sectors. In particular, we
address the electromagnetic form factors and decay constants of the
pion, kaon and D mesons. The structure of composite systems are
given by the Bethe-Salpeter (BS) amplitude of a meson formed by a
confined pair of constituent quark and antiquark,  which in our work
is written in terms of Pauli-Villars regulators. The analytical
structure contains single poles in the complex momentum space. The
BS amplitude takes into account poles due to the regulator
parameters, while the quark-antiquark cut is avoided, implying in
confined quarks with the property that the sum of the constituents
masses can be larger than the mass of the meson. The one-loop
expressions of the electroweak transition amplitudes    are
conveniently written in terms of light-front momentum. Technically,
we introduce a Wick-rotation of he minus component of the momentum
(k-minus) in the one-loop amplitudes allowing to avoid the cuts in
the complex plane of this momentum variable without crossing them.
This is particularly useful as we can study the electroweak
properties with several models of the BS amplitude with different
powers of Pauli-Villars regulators. The possibility to change the
power of the BS amplitude is interesting in order to test the
asymptotic behavior of the electromagnetic form factors searching
for a suitable form that incorporates the expected QCD decaying
power-law form. The results are compared with others models in the
literature and with the experimental data.
}

\FullConference{LIGHT CONE 2008 Relativistic Nuclear and Particle Physics\\
         July 7-11  2008\\
         Mulhouse, France}
\begin{document}

\section{Introduction}

Quantum Chromodynamics is successful in describing subatomic
processes  for  medium and large energies when short distances are
probed, and asymptotic freedom justifies the use of perturbative
amplitudes. The interesting physics that are investigated in the
electroweak structure of hadrons are in part of nonperturbative
origin. The understanding of the nonperturbative structure of
hadrons with QCD demands a lot of effort and insights from
phenomenological models that parameterizes the main properties of
confined systems fitted to data can be a helpful guidance. In this
respect, relativistic constituent models formulated in the different
forms of dynamics proposed by Dirac~\cite{Dirac49} can be used to
describe electroweak transitions. In particular the light-front (LF)
form of dynamics, is well known by its maximal number of kinematic
boosts generators and the stability of the Fock-state truncation
under such boosts. Even this limited covariance property is
essential for the calculation electroweak observables and in the
definition of the partonic content of the
hadrons~\cite{Brodskyreview}. Therefore, the description of a
covariant transition amplitude that relies on nonperturbative hadron
BS amplitudes can be decomposed in terms of matrix elements of
operators in the light-front Fock-space. In principle it is possible
to rewrite that matrix element in terms of the matrix element of an
effective operator acting in the valence sector of the hadron. Then,
a model that is able to give a LF valence wave function adequate for
describing electroweak observables can be useful phenomenological
tool to interpret results from nonperturbative calculations of QCD.

In this work, we study the electroweak properties of pseudo-scalar
mesons in the light and heavy-light sectors. In particular, we
address the electromagnetic form factors and decay constants of the
pion, kaon and D mesons. The composite systems are modeled  by
Bethe-Salpeter amplitudes written in terms of Pauli-Villars
regulators, i.e., the analytical structure contains single poles in
the complex momentum space. The mesonic Bethe-Salpeter (BS) model
amplitude takes into account poles due to the regulator parameters,
while the quark-antiquark cut is avoided, implying in confined
quarks with the property that the sum of the constituents masses can
be larger than the mass of the meson.

We adopt the light-front momentum as a tool to calculate
four-dimensional transition amplitudes involving mesons described as
confined  systems  of a constituent quark and antiquark.
Technically,  we introduce a Wick-rotation of he minus component of
the momentum $(k^-=k^0-k^3)$ in the one-loop amplitudes allowing to
avoid the cuts in the complex plane of this momentum variable
without crossing them. This is particularly useful once we can study
the electroweak properties of pseudoscalar mesons with several
models of the Bethe-Salpeter amplitude with different powers of
Pauli-Villars regulators. The possibility to change the power of the
BS amplitude is interesting for testing the asymptotic behavior of
the electromagnetic form factors and search for a suitable form that
incorporates the expected QCD decaying power-law form. The results
are compared with others models in the literature and with the
experimental data.

\section{The Model: Bethe-Salpeter Amplitude}

The model of the vertex $meson-q\bar{q}$ utilized to construct the
Bethe-Salpeter amplitude is
\begin{equation}
 \Lambda_M(k,p)= \frac{(k^2-m_1^2)  \Gamma_M ((p-k)^2-m_2^2)}
{(k^2-\lambda^2+\imath \epsilon)^n~((p-k)^2-\lambda^2_M + \imath
\epsilon)^n},
\end{equation}
where, $\lambda_M$ is the scale associated with the meson
light-front valence wave function and $n$ is the power of the
regulator. $m_1$ and $m_2$ are the quark and anti-quark masses
within the meson bound state. The factors $(k^2-m_1^2)$ and
$((p-k)^2-m_2^2)$ on the numerator of the vertex function,
$\Lambda_M(k,p)$, avoids the cuts due the $q\bar{q}$ scattering if
~$m_1+m_2$ is smaller than the meson mass. In order to the confine
the quarks, $\lambda_M$ obeys the condition $2 ~\lambda_M >~$ mass
of the meson bound state.  The model for $\Lambda_M(k,p)$ is
utilized to calculate the electromagnetic observables of
pseudoscalar mesons.

\section{One-loop electroweak transition amplitudes}

The electromagnetic form factors of pseudoscalar mesons are given by
the Mandelstam formula:
\begin{eqnarray}
\left\langle p^{\prime }\left\vert J_{q}^{\mu }\left( q^{2}\right)
\right\vert p\right\rangle  = \frac{N_{c}}{\left( 2 \pi \right)
^{4}}\int d^{4}k~Tr \left[ \Lambda_{M^{\prime }}
\left( k,p^{\prime }\right) S_{F}\left( k-p^{\prime }
\right) %\right.  \\ \nonumber &  &\left.
 J_{q}^{\mu }S(k-p)\Lambda _{M}\left( k,p\right)
S_{F}~\left( k\right) \right],
\end{eqnarray}
where $S_F(p)$ is the Feynman propagator of the quark with the
constituent mass $m_q$ and $\Lambda _{M}$ is the $meson-q\bar{q}$
vertex function presented in the last section.~$N_{c}=3$ is the
number of quark colors. $p^{\mu} $ and $p^{\prime \mu}$ are the
initial and final momenta of the system; $q^{\mu}$ is the momentum
transfer.

The pseudoscalar electromagnetic form factor is calculated with the
matrix of the electromagnetic current:
\begin{equation}
\left\langle p^{\prime }\left\vert J_{q}^{\mu }\left( q^{2}\right)
\right\vert p\right\rangle
= \left( p+p^{\prime}\right) F_{PS}^{em}(q^2)
\end{equation}
The weak decay constant of the pseudoscalar mesons is written as
\begin{equation}
\left\langle 0 \left\vert A^{\mu}(0) \right\vert p\right\rangle =
\imath \sqrt{2} f_{ps} p^{\mu},
\end{equation}
where $A^{\mu}=\bar{q}(x)\gamma^{\mu}\gamma^{5} \frac{\tau}{2}q(x)$.
The final expression for the pseudoscalar decay constant  with the
plus component of the axial-vector current, is
\begin{equation}
 \imath p^2 f_{\pi}=\frac{m}{f_{\pi}} N_c
\int \frac{dk^4}{(2 \pi)^4}
Tr\left[ \sla{p} \gamma^5  S(k) \gamma^5 S(k-p) \right] \Lambda_{M}(k,p).
\end{equation}

We use the plus component of the electromagnetic current,
~$\gamma^+=\gamma^0+\gamma^3$, to calculate the form factor of the
pseudoscalar mesons. In the next section, the Wick rotation for the
light-front approach is explained.

\section{Light-Front Wick Rotation}

The Wick rotation in the instant form quantum field theory, is
realized by the change in the component $k_0$ of the quadri-momentum
to $k_0~\longrightarrow~\imath k_0$, it is a equivalent  change to
the Euclidian space~\cite{Zuber}. The original idea was proposed by
Wick~\cite{Wick54}, where the relativistic quantum field theory
build with Minkowiski spacetime is replaced by the Euclidian space,
with the following transformation $\tau=\imath~t$, them, the
spacetime metric is writen like $ds^2=d\tau^2+dx^2+dy^2+dz^2$. The
Wick rotation is applied to solve the Bethe-Salpeter bound state
equation (BS) in the Euclidian space, because, the original BS
equation formulated in Minkowski space has singularities making it
difficult to solve. The idea is to use the Wick rotation to avoid
the singularities in the Bethe-Salpeter equation by performing the
calculation in the Euclidian space.

In the light-front quantum field theory, the choice of frame used to
calculated Feynman amplitudes is helpful, because the analytical
structure of the vertex function of the bound state-$q\bar{q}$ pole
position depends on the energy integration $k^-$.

The idea of the Wick rotation within the light-front approach comes
to avoid the problems related with the poles of the BS amplitudes
and the frame used to calculate the Feynman amplitudes. The choice
of a frame for the computation of a given amplitude within
light-front quantum field theory is sensible because it is related
to the breaking of covariance by the valence terms and the
contribution non-valence components of the light-front wave
function~\cite{demelo97,Naus98,demelo2002}.

A common frame used to compute form factors in the light-front is
the Breit-frame, characterized by a zero plus component of the
momentum transfer, $q^+=0$. In the past, this frame was believed
free of the pair terms contributions or zero modes, but is not
really true. Not only the frame dependence is important in the
light-front approach, but also which component of the
electromagnetic current is used to extract the observables.  This
was evident after the works done in the
references~\cite{demelo99,Ji2001}. For example for the pion case,
the electromagnetic form factor was calculated in the same model for
the plus and minus component of the electromagnetic current, and
compared with the equal time calculation. The minus component of the
electromagnetic current, besides the valence contribution, has also
the non-valence contribution,  then, the full covariance of the
minus component of the electromagnetic current needed this two
contributions, valence and nonvalence~\cite{demelo99}. However, the
plus component of the electromagnetic current do not have other
contributions besides the valence one, and the calculated matrix
elements of the electromagnetic current is exactly the same one
obtained in  an equal time calculation or instant form
approach~\cite{demelo99}.

The use of the Wick rotation with light-front formalism is explored
in the next section, where the electromagnetic form factors and weak
decay constants are calculated.

\vskip 0.78cm
\begin{figure}[h]
\centerline{
\begin{picture}(330,130)(0,0)
\LongArrow(-10,52)(220,52)
\LongArrow(100,-13)(100,150)
%% Cortorno // Integration
\Line(19,14)(182,90)
\ArrowLine(100,52)(180,52)
\ArrowLine(90,52)(10,52)
\ArrowArc(120,40)(80,8.5,39)  %% arc up
\ArrowArc(30,42)(30,160,248.5)   %% arc down
\put(90,60){\makebox(0,0)[br]{$k^-_{2}$}}
\Vertex(86,78){2}
\put(300,130){\makebox(0,0)[br]{{\bf  \ $k^-$ \ \ $\Longrightarrow$
\  \  \ $k^-e^{\imath \theta}$ ;  \ \ \ \  $0^0 < \theta < 90^0 $}}}
\Vertex(56,78){2}
\put(60,60){\makebox(0,0)[br]{$k^-_{1}$}}
\put(150,10){\makebox(0,0)[br]{$k^-_{3}$}}
\Vertex(140,30){2}
\Vertex(170,30){2}
\put(180,10){\makebox(0,0)[br]{$k^-_{4}$}}
\end{picture}
}
\caption{Wick rotation in the light-front coordinates}
\label{fig4.2}
\end{figure}
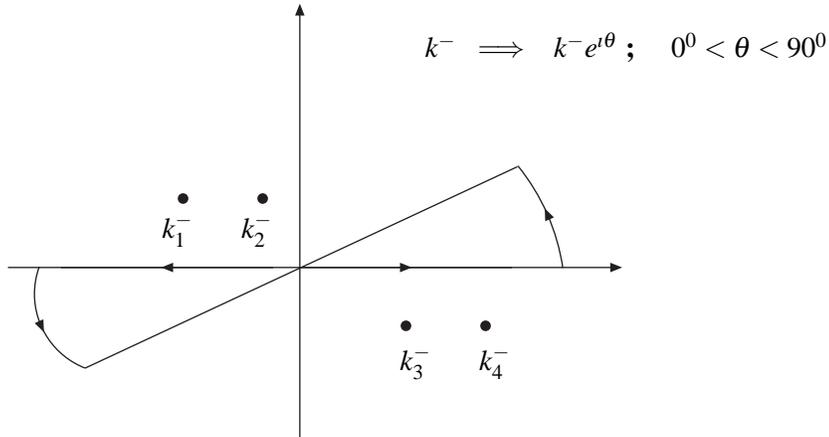
%%%%\end{large}

\vspace{-.70003cm}

\section{Results and Conclusion}

We have calculated with the presented model the electromagnetic form
factors and the weak decay constants for the light pseudoscalar
mesons, pion, kaon and $D^+$ with the Wick rotation performed within
the light-front approach.  The present model, was compared with
other light-front models and dispersion relation
calculations~\cite{demelo2002,demelo99,BJP2008}.

The table 1, shows the results for the electromagnetic radius and
weak decay constants for the pion and kaon, compared with the
experimental data. The parameters of the model are, the constituents
quark masses, the power in the vertex, $n$, and  $\lambda_M$. The
value of the masses $\lambda_M$ are given by the fit of the
experimental value of the weak decay constants $f_{ps}$. The masses
of the constituents quarks used are $m_u=0.220$~GeV and
$m_s=0.508$~GeV.
 The value of the electromagnetic radius of the pion is
aproximately $14\%$ below of the experimental value
($0.675\pm0.02$)~\cite{Amendolia86} with n=2 and the mass scale is
$\lambda_M=0.542~GeV$ for the fits the pion decay constant is
$f_{\pi}=92.4~MeV$. The electromagnetic form factor for the pion is
presented in figures 1 and 2; and the kaon an $D^+$ electromagnetic
form factor are show in figure 3.

In conclusion, the electromagnetic form factors and weak decay
constants calculated with the present light-front model were
compared with others models and reproduced very well the
experimental data.

\begin{center}
\begin{table}[t,b,h]
\begin{tabular}{|l|l|l|l|l|}
\hline
%%Light-Front Covariant Model
\hline
   & n & $r_{PS}(fm)$ & $f_{PS}(MeV)$ &
$\lambda_M~(MeV)$ \\ %%$M_R(MeV)$ \\
\hline
 Pion ~(139~MeV)   & ~2     &  ~0.576    & ~92.4   & ~542 \\
$m_u=220~(MeV)$  & ~3   & ~0.494   & ~92.4    &  ~926 \\
 $m_d=220~(MeV)$    &  ~4   & ~0.456  & ~92.4 & ~1255 \\
\hline
Exp.~~(Pion)           &  &  ~0.672  & ~92.42 &  \\
\hline
Kaon~~(494~MeV) & ~2 & ~0.474   & ~113 & ~648  \\
$m_s=0.508~(MeV) $ & ~3 & ~0.453  & ~113 & ~933 \\
$m_u=0.220~(MeV)$ & ~4 &  ~0.450 & ~113 & ~1156  \\
\hline
Exp.(Kaon)   &   & ~0.560 & ~113 &  ~  \\
\hline
\hline
\end{tabular} \\
\caption {Results for the pseudoscalar weak decay constants
$f_{ps}$ and the electromagnetic radius for the pion and kaon
compared with the experimental data.
}
\label{table1}
\end{table}
\end{center}

\begin{figure}[tbh!]
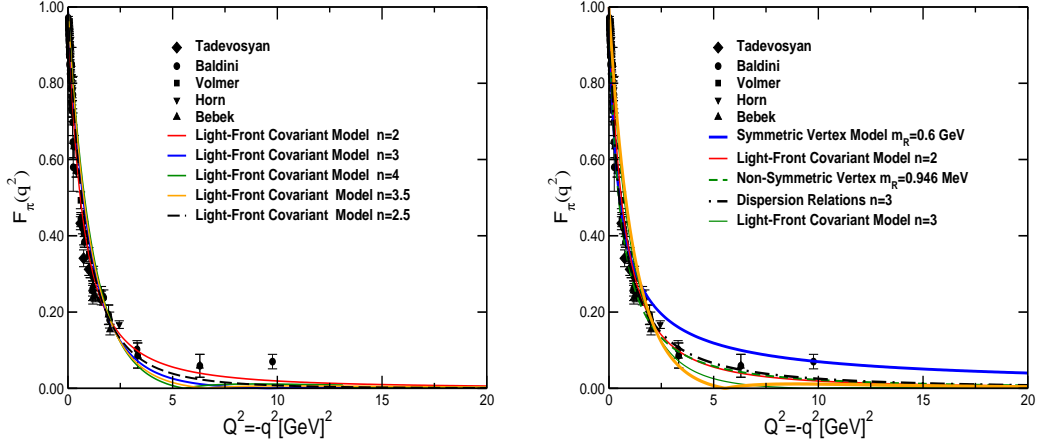

%%\vspace{-0.2003cm}
\centerline{\epsfig{figure=lfwcovpion.eps,width=6.5cm,height=6.5cm}
\hspace{0.5cm} \epsfig{figure=lfwpion.eps,width=6.5cm,height=6.5cm}}
\caption{ The left figure shows the pion electromagnetic form factor
calculated with the light-front covariant model for different n
values and it is compared to the experimental data.  In the right
figure, the light-front covariant model is compared with other
light-front models from references~\cite{demelo2002,demelo99} and
dispersion relations~\cite{BJP2008}, also with the experimental
data~\cite{piondata}. } \label{fig1}
\end{figure}

\begin{figure}[tbh!]
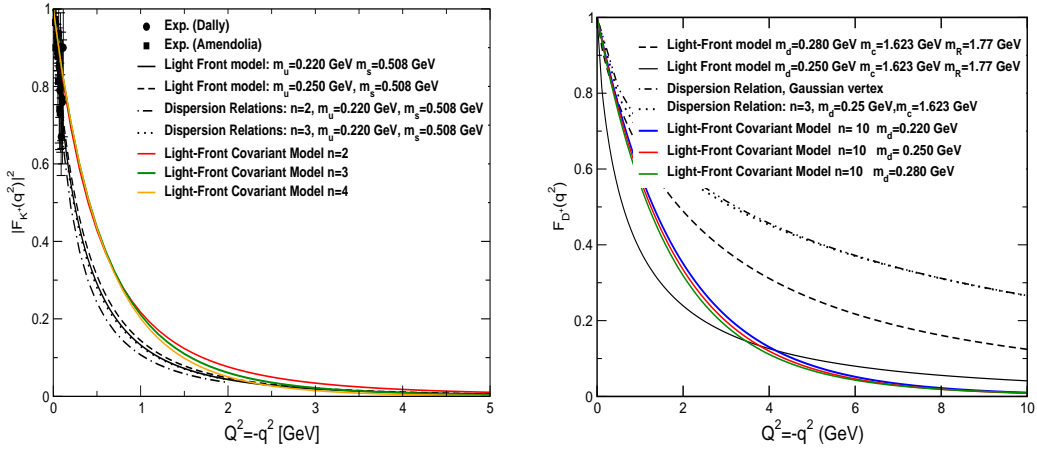

%%\vspace{-0.2218cm}
\centerline{\epsfig{figure=lfwkaon.eps,width=6.5cm,height=6.5cm}
\hspace{0.5cm}
\epsfig{figure=figdplus81.eps,width=6.5cm,height=6.5cm}} 
\caption{The kaon electromagnetic form factor (left) calculated with light-front
covariant model and other models is  compared with the experimental
data ~\cite{Amendolia86,kaondata}. The eletromagnetic form factor for 
the $D^+$ ~meson with the covariant light-front model and compared with others models. 
}
\label{fig2}
\end{figure}

\end{document}